\documentclass[aps,pra,reprint,superscriptaddress,nofootinbib]{revtex4-2}
\usepackage{amsmath,amssymb,amsthm,braket}
\usepackage{graphicx} 
\usepackage{braket}
\usepackage{diagbox} 
\usepackage{color}

\graphicspath{ {./} }

\def\sket#1{|#1\,\rangle\hspace{-2mm}\rangle}
\def\sbra#1{\langle\hspace{-2mm}\langle\,#1|}
\def\sbraket#1#2{\langle\hspace{-2mm}\langle\,#1|#2\,\rangle\hspace{-2mm}\rangle}

\theoremstyle{plain}
\newtheorem{thm}{Theorem}
\newtheorem{cor}{Corollary}

\begin{document}
\title{Quantum channel decomposition with pre- and post-selection}

\author{Ryo Nagai}
\email{ryo.nagai.jd@hitachi.com}
\affiliation{
Center for Exploratory Research, 
Research and Development Group, 
Hitachi, Ltd., 
Kokubunji, Tokyo, 185-8601, Japan.}
\affiliation{Quantum Computing Center, Keio University, 3-14-1 Hiyoshi, Kohoku-ku, Yokohama, Kanagawa, 223-8522, Japan}

\author{Shu Kanno}
\affiliation{Mitsubishi Chemical
Corporation, Science \& Innovation
Center, Yokohama, 227-8502, Japan}
\affiliation{Quantum Computing Center, Keio University, 3-14-1 Hiyoshi, Kohoku-ku, Yokohama, Kanagawa, 223-8522, Japan}

\author{Yuki Sato}
\affiliation{Toyota Central R\&D Labs., Inc., 1-4-14 Koraku, Bunkyo-ku, Tokyo, 112-0004, Japan}
\affiliation{Quantum Computing Center, Keio University, 3-14-1 Hiyoshi, Kohoku-ku, Yokohama, Kanagawa, 223-8522, Japan}

\author{Naoki Yamamoto}
\affiliation{Quantum Computing Center, Keio University, 3-14-1 Hiyoshi, Kohoku-ku, Yokohama, Kanagawa, 223-8522, Japan}
\affiliation{Department of Applied Physics and Physico-Informatics, Keio University, 3-14-1 Hiyoshi, Kohoku-ku, Yokohama 223-8522, Japan}

\date{\today}

\begin{abstract}

The quantum channel decomposition techniques, which contain the so-called 
probabilistic error cancellation and gate/wire cutting, are powerful approach 
for simulating a hard-to-implement (or an ideal) unitary operation by 
concurrently executing relatively easy-to-implement (or noisy) quantum channels. 
However, such virtual simulation necessitates an exponentially large number of 
decompositions, thereby significantly limiting their practical applicability. 
This paper proposes a channel decomposition method for target unitaries that 
have their input and output conditioned on specific quantum states, namely 
unitaries with pre- and post-selection. 
Specifically, we explicitly determine the requisite number of decomposing channels, 
which could be significantly smaller than the selection-free scenario. 
Furthermore, we elucidate the structure of the resulting decomposed unitary. 
We demonstrate an application of this approach to the quantum linear solver 
algorithm, highlighting the efficacy of the proposed method. 

\end{abstract}

\maketitle

\section{Introduction}

Running a large quantum circuit is a challenging task, unless the ideal 
quantum error correction would be incorporated. 
To simulate such circuit without error correction or to relax the strict condition 
for realizing error correction, several techniques have been developed -- the 
quantum channel decomposition is one such effective method for this purpose.

The idea of channel decomposition is, as the name implies, to decompose 
a hard-to-implement (or an ideal) quantum channel to the sum of relatively 
easy-to-implement (or realistic) channels. 
The so-called probabilistic 
error cancellation \cite{temme2017error, endo2018practical,takagi2021optimal,piveteau2022quasiprobability} can 
be formulated in this framework. 
Also, especially when the target channel is an ideal unitary operator and the 
decomposing channels are given by products of smaller unitaries, then the method 
is called the gate cutting 
\cite{bravyi2016trading,mitarai2021constructing,mitarai2021overhead,
marshall2022high,piveteau2022circuit,takeuchi2022divide,bechtold2023investigating}. 
In addition, when the target gate is a set of identity operators along 
the lines of a quantum circuit diagram, it is called 
the wire cutting \cite{peng2020simulating,perlin2021quantum,ayral2020quantum,tang2021cutqc,
ying2023experimental,tang2022scaleqc,uchehara2022rotation,chen2022approximate,takeuchi2022divide,
lowe2023fast,ufrecht2023cutting,brenner2023optimal,harada2023optimal,pednault2023alternative,chen2023efficient}. 
However, these channel decomposition methods have the common issue of 
exponential computational overhead; that is, the number of decomposing channel 
grows exponentially with respect to the number of qubits of the target channel.

In this paper, we propose a general channel decomposition method that reduces 
the number of decomposition, though the exponential scaling is kept. 
The idea is to pre- and post-select the states for a target unitary gate. 
More precisely, the input state to the target unitary gate and the output state 
from the target are both limited to certain type of quantum states living 
in subspaces of the entire Hilbert space. 
This situation is often found in several quantum algorithms. 
Actually, when the target is a unitary operator representing the entire circuit, 
the initial state is usually fixed (typically, to the product of $\ket{0}$); 
also, when the target unitary gate corresponds to a sub-circuit of the entire 
circuit, the input state to that unitary can still be restricted, as in the case 
of Grover search algorithm \cite{grover1996fast}. 
Also, the output state is confined to a subspace in some quantum algorithms, 
such as probabilistic state preparation algorithms that succeed only when a part 
of the output state is projected to a certain fixed state by measurement; 
this idea is applicable to both cases where the target unitary gate is an entire 
circuit and a part of the entire circuit. 
Under this pre- and post-selection condition, we explicitly identify the 
necessary number of decomposing channels (Theorem~1) and moreover the structure 
of the resulting decomposed unitary gate (Theorem~2). 
Then, we demonstrate the proposed method with a 5-qubits quantum linear solver 
algorithm (Harrow-Hassidim-Lloyd, HHL algorithm) \cite{harrow2009quantum}; notably, 
while the original circuit is of 
212 depth and contains 108 CNOT gates, it can be decomposed to only three 
circuits that are composed of the product of one-qubit gates and thus much 
easier to implement than the original.

\section{Preliminaries}
\subsection{Quantum channel decomposition}

We consider the following $n$-qubit quantum channel:
\begin{align}
\rho 
\mapsto
\rho'
=
U^{(n)} \rho\,U^{(n)\dag}
\,.
\label{eq:linmap}
\end{align}
Here $U^{(n)}$ is an $N\times N$ unitary matrix, where $N=2^n$. 
Also, $\rho$ and $\rho'$ denote the density matrix of the $n$-qubit 
system before and after the unitary gate operation, respectively. 
For later convenience, we define the linear map $[U]$ as 
\begin{align}
[U]\rho
:=
U\rho \,U^\dag
\,.
\label{eq:supopU}
\end{align}
Using this symbol, we can express Eq.~\eqref{eq:linmap} as 
\begin{align}
\rho'=[U^{(n)}]\rho 
\,.
\label{eq:linmap-2}
\end{align}

One can 
decompose $[U^{(n)}]$ into a set of the linearly 
independent completely positive and trace-nonincreasing maps \cite{endo2018practical,mitarai2021constructing,mitarai2021overhead,takagi2021optimal}
$\{[V^{(n)}_i]\}$ as 
\begin{align}
    [U^{(n)}]
    \,=\,
    \sum_{i=1}^D 
    c_i \,
    [V^{(n)}_i]
    \,,
    \label{eq:qcd}
\end{align}
where the coefficients $\{c_i\}$ are real.
$\{[V^{(n)}_i]\}$ is the set of the quantum channels which are easier to implement than $[U^{(n)}]$. 
We will specify $\{[V^{(n)}_i]\}$ later. 
The number of the independent maps,
$D$, should be larger than $N^4=16^n$ to satisfy the equality in 
Eq.~(\ref{eq:qcd}) for a general $[U^{(n)}]$.
The coefficients $\{c_i\}$ are uniquely determined when $D=16^n$,
while they are ambiguously determined when $D>16^n$.
We call the decomposition (\ref{eq:qcd}) the ``quantum channel decomposition''
{\footnote{
We can also decompose the intermediate quantum channel in an entire quantum 
circuit. 
Let us consider the $n$-qubit quantum channel $[U^{(n)}]$ 
consisting of a sequence of $m$ quantum channels, 
$[U^{(n)}]=[U^{(n)}_m \cdots U^{(n)}_1]$. 
Since the linear map $[U^{(n)}]$ can be regarded as the composite map given as 
$[U^{(n)}_m]\circ\cdots\circ [U^{(n)}_1]$, we can consider the decomposing problem 
of only the $k$-th channel $[U^{(n)}_k]$ $(1\leq k \leq m)$.
}}.

The motivation to conduct the channel decomposition \eqref{eq:qcd} arises when 
$[U^{(n)}]$ is more difficult to implement than $\{[V^{(n)}_i]\}$. 
Actually, Eq.~(\ref{eq:qcd}) implies that we can simulate the output of the harder 
quantum channel $[U^{(n)}]$ by combining the output of the easier quantum channels 
$\{[V^{(n)}_i]\}$. 
In particular, the Monte-Carlo method can be applied to calculate the expectation 
of an observable $B$ for the output state $\rho'$ via 
\begin{align}
     {\rm Tr}\left(B [U^{(n)}]\rho\right)
     = \gamma \sum_{i=1}^D \frac{|c_i|}{\gamma} \, {\rm sign}(c_i) 
         \, {\rm Tr}\left(B [V_i^{(n)}]\rho\right), 
\end{align}
where $\gamma=\sum_{i=1}^D|c_i|$. 
That is, by sampling $\{[V^{(n)}_i]\}$ with probability $|c_i|/\gamma$ 
followed by calculating the expectation of $B$, the hard calculation of the 
expectation of $B$ becomes doable. 
The overhead for this simulation is quantified by $\gamma$, 
because the approximation error of the expectation is 
$\mathcal{O}(\gamma/\sqrt{N_{\rm{s}}})$ with $N_{\rm{s}}$ being the total 
number of sampling.
Hence, we call $\gamma$ the sampling overhead.

\subsection{Gate decomposition}

The goal of gate decomposition is to decompose the quantum channel $[U^{(n)}]$ 
into a set of quantum channels consisting of only local quantum operations 
as follows: 
\begin{align}
    [U^{(n)}]
    \,=\,
    \sum_{i=1}^D 
    c_i\, 
    [V^{(1)}_{i_1} 
    \otimes V^{(1)}_{i_2}
    \otimes \cdots
    \otimes V^{(1)}_{i_n}]
    \,,
    \label{eq:gate-cutting}
\end{align}
where $i_k$ runs from $1$ to $D_k\,(\geq 16)$ and $i$ runs in the form 
$i=\sum_{k=1}^n(\prod_{k'=1}^k D_{k'-1})i_k$ with $D_0=1$. 
$D$ is given as $D=D_1D_2\cdots D_n$.
$\{V^{(1)}_{i_a}\,(a=1,2,\cdots,n)\}$ denotes a set of one-qubit operations. 
The one-qubit operations consist of not only one-qubit unitary gate but also 
non-unitary operations such as measurement operations.

Note that the basis $\{V^{(1)}_{i_a}\}$ is not unique. 
Also, the coefficients $\{c_i\}$ depend on the basis.
In this paper, we employ the basis introduced in Ref.~\cite{endo2018practical}
which is listed in Table \ref{tab:basis}.
Other bases are considered by Refs.~\cite{mitarai2021constructing,mitarai2021overhead,takagi2021optimal}, 
for examples.

\begingroup
\squeezetable
\begin{table}
    \caption{
    16 linearly independent operations for 
    the one-qubit gate decomposition.
    $I$ denotes $2\times 2$ identity matrix.
    $X$, $Y$ and $Z$ are Pauli matrices. 
    This set of operators was introduced in Ref.~\cite{endo2018practical}. 
   }
    \label{tab:basis}
    \centering
    \vspace{0.2cm}
    {\renewcommand\arraystretch{1.5}
    \begin{tabular}{c|c}
      \hline
     $[{V}^{(1)}_1]$ & \qquad $[{I}]$ \\
     $[{V}^{(1)}_2]$ & \qquad $[{X}]$ \\
     $[{V}^{(1)}_3]$ & \qquad $[{Y}]$ \\
     $[{V}^{(1)}_4]$ & \qquad $[{Z}]$ \\
     $[{V}^{(1)}_5]$ & \qquad $[{R}_X]=[(1/\sqrt{2})({I}+i{X})]$ \\
     $[{V}^{(1)}_6]$ & \qquad $[{R}_Y]=[(1/\sqrt{2})({I}+i{Y})]$ \\
     $[{V}^{(1)}_7]$ & \qquad $[{R}_Z]=[(1/\sqrt{2})({I}+i{Z})]$ \\
     $[{V}^{(1)}_8]$ & \qquad $[{R}_{YZ}]=[(1/\sqrt{2})({Y}+{Z})]$ \\
     $[{V}^{(1)}_9]$ & \qquad $[{R}_{ZX}]=[(1/\sqrt{2})({Z}+{X})]$ \\
     $[{V}^{(1)}_{10}]$ & \qquad $[{R}_{XY}]=[(1/\sqrt{2})({X}+{Y})]$ \\
     $[{V}^{(1)}_{11}]$ & \qquad $[{\pi}_X]=[(1/2)({I}+{X})]$  \\
     $[{V}^{(1)}_{12}]$ & \qquad $[{\pi}_Y]=[(1/2)({I}+{Y})]$ \\
     $[{V}^{(1)}_{13}]$ & \qquad $[{\pi}_Z]=[(1/2)({I}+{Z})]$   \\
     $[{V}^{(1)}_{14}]$ & \qquad $[{\pi}_{YZ}]=[(1/2)({Y}+i{Z})]$  \\
     $[{V}^{(1)}_{15}]$ & \qquad $[{\pi}_{ZX}]=[(1/2)({Z}+i{X})]$  \\
     $[{V}^{(1)}_{16}]$ & \qquad $[{\pi}_{XY}]=[(1/2)({X}+i{Y})]$  \\
      \hline
    \end{tabular}
    }
\end{table}
\endgroup

Once the basis is fixed, we can 
determine the coefficients $\{c_i\}$ 
by solving Eq.~(\ref{eq:gate-cutting}) for a given $U^{(n)}$. 
Here we present some results of the gate decomposition of specific targets as 
examples. 
We consider controlled-NOT (CNOT), Toffoli, and three-qubit Quantum Fourier 
Transformation (QFT$_3$) gates as the target unitary channel. 
In the computational basis, these unitary matrices are represented by
\begin{align}
\mbox{CNOT}:
\quad
U^{(n)}
&\,=\,
\begin{pmatrix}
    1 & 0 & 0 & 0 \\
    0 & 1 & 0 & 0 \\
    0 & 0 & 0 & 1 \\
    0 & 0 & 1 & 0 
\end{pmatrix}
\,,
\label{eq:CNOT}\\
\mbox{Toffoli}:
\quad
U^{(n)}
&\,=\,
\begin{pmatrix}
    1 & 0 & 0 & 0 & 0 & 0 & 0 & 0 \\
    0 & 1 & 0 & 0 & 0 & 0 & 0 & 0 \\
    0 & 0 & 1 & 0 & 0 & 0 & 0 & 0 \\
    0 & 0 & 0 & 1 & 0 & 0 & 0 & 0 \\
    0 & 0 & 0 & 0 & 1 & 0 & 0 & 0 \\
    0 & 0 & 0 & 0 & 0 & 1 & 0 & 0 \\
    0 & 0 & 0 & 0 & 0 & 0 & 0 & 1 \\
    0 & 0 & 0 & 0 & 0 & 0 & 1 & 0 
\end{pmatrix}
\,,
\label{eq:Toffoli}\\
\mbox{QFT}_{3}:
\quad
U^{(n)}
&\,=\,
\frac{1}{\sqrt{8}}
\begin{pmatrix}
    1 & 1 & 1 & 1 & 1 & 1 & 1 & 1 \\
    1 & \omega & \omega^2 & \omega^3 & \omega^4 & \omega^5 & \omega^6 & \omega^7 \\
    1 & \omega^2 & \omega^4 & \omega^6 & 1      & \omega^2 & \omega^4 & \omega^6 \\
    1 & \omega^3 & \omega^6 & \omega & \omega^4 & \omega^7 & \omega^2 & \omega^5 \\
    1 & \omega^4 & 1        & \omega^4 & 1      & \omega^4 & 1      & \omega^4 \\
    1 & \omega^5 & \omega^2 & \omega^7 & \omega^4 & \omega^1 & \omega^6 & \omega^3 \\
    1 & \omega^6 & \omega^4 & \omega^2 & 1      & \omega^6 & \omega^4 & \omega^2 \\
    1 & \omega^7 & \omega^6 & \omega^5 & \omega^4 & \omega^3 & \omega^2 & \omega 
\end{pmatrix}
\,,
\label{eq:QFT}
\end{align}
with $\omega=e^{i\pi/4}$. 
The sampling overheads $\gamma$ for the gate decomposition of these unitary 
channels $[U^{(n)}]$ are obtained as
\begin{align}
\mbox{CNOT}:
\quad
&
\gamma
=
\sum_{i=1}^{16^2}|c_i|
=9
\,,\\
\mbox{Toffoli}:
\quad
&
\gamma
=
\sum_{i=1}^{16^3}|c_i|
=37
\,,\\
\mbox{QFT}_{3}:
\quad
&
\gamma
=
\sum_{i=1}^{16^3}|c_i|
\simeq 261
\,.
\end{align}
The explicit expressions of the gate decomposition for the above examples 
are summarized in Appendix \ref{app:gate-decomp}.

\section{Main results}

\subsection{Notation}

Let $A$ be an $m\times n$ matrix. 
We define the vectorization of $A$, denoted by $\sket{A}$, as the 
$(mn\times 1)$-vector with the columns of the matrix 
$A$ stacked on top of each other \cite{horn2012matrix}. 
We also define $\sbra{A}$ as $\sbra{A}=(\sket{A})^\dag$. 
The inner product $\sbraket{A}{B}$ is calculated as 
$\sbraket{A}{B}=\mbox{Tr}[A^\dag B]$. 
We have
\begin{align}
\sket{ABC}
\,=\,
(C^T\otimes A)\sket{B}
\,.
\label{eq:sketABC}
\end{align}
for arbitrary matrices $A$, $B$ and $C$ \cite{horn2012matrix}. 
We will use this equation later.

Given an $n$-qubit quantum channel $\mathcal{A}$,
we introduce the Choi-matrix $\mbox{C}(\mathcal{A})$ as
\begin{align}
\mbox{C}(\mathcal{A})
\,=\,
\sum_{i=1}^{N}
\sum_{j=1}^{N}
E_{ij}
\otimes 
\mathcal{A}
(E_{ij})
\,,
\end{align}
where $N=2^n$ and 
$E_{ij}=\ket{i}\!\!\bra{j}$ in the computational basis.
In particular, for the quantum channel 
$\mathcal{A}(\rho)=\sum_{i=1}^a c_i [V^{(n)}_i]\rho$, 
we have
\begin{align}
\mbox{C}(\mathcal{A})
=
\sum_{i=1}^a c_i 
\sket{V^{(n)}_i}~\sbra{V^{(n)}_i}
\,.
\label{eq:defChoi}
\end{align}
It is known that the mapping $\mathcal{A}\mapsto \mbox{C}(\mathcal{A})$ is 
isomorphism and affine (Jamiolkowski–Choi isomorphism) \cite{de1967linear,jamiolkowski1972linear,choi1975completely,bergholm2015diagrammatic}.

\subsection{Quantum channel with pre- and post-selection}

We consider the $n$-qubit quantum channel $[U^{(n)}]$ given in 
Eq.~(\ref{eq:linmap-2}), where the input state $\rho$ and the output state 
$\rho'$ 
are specified somehow. 
In particular, we are often interested in the situation where these 
states are restricted in some subspaces; such input and output states, $\rho_{\mathrm{in}}$ and $\rho_{\mathrm{out}}$, can be expressed as 
\begin{align}
    \rho_{\mathrm{in}} 
    \,=\, 
    \frac{P_{\mathrm{in}}\rho P_{\mathrm{in}}}
    {\mbox{Tr}(\rho P_{\mathrm{in}})}
    \,,\quad
    \rho_{\mathrm{out}} 
    \,=\, 
    \frac{P_{\mathrm{out}}\rho' P_{\mathrm{out}}}
    {\mbox{Tr}(\rho' P_{\mathrm{out}})}
    \,.
\end{align}
Here, $P_{\mathrm{in}}$ and $P_{\mathrm{out}}$ are the projection matrices 
(Recall ${P^\dag_{\mathrm{in,out}}=P_{\mathrm{in,out}}}$ and 
$P^2_{\mathrm{in,out}}=P_{\mathrm{in,out}}$) which respectively describe 
the pre- and post-selection of the quantum states. 
We note that $\max(r_{\mathrm{in}},r_{\mathrm{out}})\leq N=2^n$
where
\begin{align}
    r_{\mathrm{in}}:= {\mathrm{rank}}P_{\mathrm{in}}
    \,,
    \quad
    r_{\mathrm{out}}:= {\mathrm{rank}}P_{\mathrm{out}}
    \,.
\end{align}

The quantum channel with the post- and pre-selection 
is described as
\begin{align}
    [P_{\mathrm{out}}U^{(n)}P_{\mathrm{in}}]
    \,,
    \label{eq:qcwps}
\end{align}
up to the normalization factor $\mbox{Tr}(\rho P_{\mathrm{in}})$ 
and $\mbox{Tr}(\rho' P_{\mathrm{out}})$. 
Now, $[P_{\mathrm{out}}U^{(n)}P_{\mathrm{in}}]$ can be decomposed with 
fewer bases than the case we decompose $[U^{(n)}]$. 
The reason is as follows. 
The map $[P_{\mathrm{out}}U^{(n)}P_{\mathrm{in}}]$ except for the parts 
determined by the projection operators is described by the linear map 
from the quantum states on $r_{\mathrm{in}}$-dimensional Hilbert space to 
those on $r_{\mathrm{out}}$-dimensional Hilbert space. 
This means that the quantum channel with post- and pre-selection
$[P_{\mathrm{out}}U^{(n)}P_{\mathrm{in}}]$ is inherently described by 
$r^2_{\mathrm{in}}\times r^2_{\mathrm{out}}$ matrix.
The size of the matrix representation of 
$[P_{\mathrm{out}}U^{(n)}P_{\mathrm{in}}]$ is therefore smaller than 
that of the matrix representation of $[U^{(n)}]$, $N^4=16^n$.

To be more concrete, let us discuss the structure of 
$[P_{\rm{out}}U^{(n)}P_{\rm{in}}]$ explicitly. 
The Choi-matrix of the quantum channel with pre- and post-selection 
(\ref{eq:qcwps}) is given as
\begin{align}
\mbox{C}([P_{\mathrm{out}}U^{(n)}P_{\mathrm{in}}])
\,=\,
{\bf{P}}
\mbox{C}([U^{(n)}])
{\bf{P}}
\,,
\label{eq:ChoiPUP}
\end{align}
where ${\bf{P}}\,=\,
P_{\mathrm{in}}
\otimes 
P_{\mathrm{out}}$.
Here we used Eqs.~(\ref{eq:sketABC}) and (\ref{eq:defChoi}).
The matrix structure of the right-handed side of Eq.~(\ref{eq:ChoiPUP}) 
can be more clarified by diagonalizing the projection matrices as
\begin{align}
    P_{\mathrm{in}}
    &\,=\,
    O^\dagger_{\mathrm{in}}
    P_{d,\mathrm{in}}
    O_{\mathrm{in}}
    \,,
    \quad
    P_{d,\mathrm{in}}
    \,=\,
    \begin{pmatrix}
     I_{r_{\mathrm{in}}} &  \phantom{I_a} \\
     \phantom{I_a} & \phantom{I_a} 
    \end{pmatrix}
    \,,
    \label{eq:Pin}
    \\
    P_{\mathrm{out}}
    &\,=\,
    O_{\mathrm{out}}
    P_{d,\mathrm{out}}
    O^\dagger_{\mathrm{out}}
    \,,
    \quad
    P_{d,\mathrm{out}}
    \,=\,
    \begin{pmatrix}
     I_{r_{\mathrm{out}}} &  \phantom{I_a} \\
     \phantom{I_a} & \phantom{I_a}  
    \end{pmatrix}
    \,,
    \label{eq:Pout}
\end{align}
where $I_m$ denotes the $m\times m$ identity matrix.
The components of the blank part are zero.
$O_{\mathrm{in}}$ and $O_{\mathrm{out}}$ are 
known
$N\times N$ unitary 
matrices that are determined to satisfy Eqs.~(\ref{eq:Pin}) and (\ref{eq:Pout}), 
respectively.
We here employ the computational basis for the above matrix representations just for simplicity.

We comment on the implementation of the diagonalization unitary matrices, 
$O_{\mathrm{in}}$ and $O_{\mathrm{out}}$, on a circuit. 
When the projection operator can be written as $P=\ket{\psi}\!\!\bra{\psi}$ 
with $\ket{\psi}$ being a pure state, the unitary matrix can be implemented 
as the corresponding state preparation operator. 
In addition, by applying the purification technique, we can implement the 
unitary operator as a state preparation process, for the general case where 
$P$ has a form of mixed state, i.e., a linear combination of the pure projection 
operators.

Plugging Eqs.~(\ref{eq:Pin}) and (\ref{eq:Pout}) into Eq.~(\ref{eq:ChoiPUP})
and using Eq.~(\ref{eq:sketABC}), we obtain
\begin{align}
\mbox{C}([P_{\mathrm{out}}U^{(n)}P_{\mathrm{in}}])
&\,=\,
{\bf{O}}^\dagger
{\bf{P}}_d
\mbox{C}([O_{\mathrm{out}}U^{(n)}O_{\mathrm{in}}])
{\bf{P}}_d
{\bf{O}}
\,,
\label{eq:CPUP_2}
\end{align}
where ${\bf{O}}
=
O_{\mathrm{in}}\otimes O^\dagger_{\mathrm{out}}$
and 
\begin{align}
{\bf{P}}_d
=
P_{d,\mathrm{in}}\otimes P_{d,\mathrm{out}}
\,.
\end{align}
It should be noted that ${\bf{P}}_d
\mbox{C}([O_{\mathrm{out}}U^{(n)}O_{\mathrm{in}}])
{\bf{P}}_d$ is the principal submatrix of $\mbox{C}([O_{\mathrm{out}}U^{(n)}O_{\mathrm{in}}])$.
The submatrix can be represented as 
\begin{align}
{\bf{P}}_d
\mbox{C}([O_{\mathrm{out}}U^{(n)}O_{\mathrm{in}}])
{\bf{P}}_d
\,=\,
\sket{V^{(n)}}
~
\sbra{V^{(n)}}
\,,
\label{eq:PCP}
\end{align}
where $V^{(n)}$ is the $N\times N$ matrix of the form 
\begin{align}
\label{eq:V1}
V^{(n)}
\,=\,
\begin{pmatrix}
V & \phantom{V}\\
\phantom{V} & \phantom{V}
\end{pmatrix}.
\end{align}
Here, $V$ is the $r_{\mathrm{out}}\times r_{\mathrm{in}}$ matrix given by
\begin{align}
V
\,=\,
\begin{pmatrix}
(v_1)_1 & \cdots & (v_1)_{r_{\mathrm{in}}} \\ 
\vdots & \ddots & \vdots \\
(v_{r_{\mathrm{out}}})_1 & \cdots & (v_{r_{\mathrm{out}}})_{r_{\mathrm{in}}} \\ 
\end{pmatrix}
\end{align}
with
\begin{align}
(v_i)_{j}
&\,=\,
\sbraket{E_{ji}}{O_{\mathrm{out}}U^{(n)}O_{\mathrm{in}}}
\nonumber\\
&\,=\,
\sbraket{O^\dagger_{\mathrm{out}}E_{ji} O^\dagger_{\mathrm{in}}}{U^{(n)}}
\,.
\label{eq:V2}
\end{align}
Combining Eqs.~(\ref{eq:CPUP_2}) and (\ref{eq:PCP}),
we finally arrive at $\mbox{C}([P_{\mathrm{out}}U^{(n)}P_{\mathrm{in}}])
\,=\,
\mbox{C}([O_{\mathrm{out}}V^{(n)}O_{\mathrm{in}}])$
which results in
\begin{align}
[P_{\mathrm{out}}U^{(n)}P_{\mathrm{in}}]
\,=\,
[O_{\mathrm{out}}V^{(n)}O_{\mathrm{in}}]
\,.
\label{eq:PUP}
\end{align}
It is now clear that the $U^{(n)}$-dependent part of  
$[P_{\mathrm{out}}U^{(n)}P_{\mathrm{in}}]$ is summarized into $V^{(n)}$ 
whose non-trivial components are collected in the 
$r_{\rm{out}}\times r_{\rm{in}}$ matrix $V$.

How can $V^{(n)}$ be implemented in a quantum circuit, then?
We express $V^{(n)}$ givn in Eq.~(\ref{eq:V1}) as 
\begin{align}
V^{(n)}
\,=\,
\ket{0}\!\!\bra{0}^{\otimes (n-\tilde{n})}\otimes \widetilde{V}^{(\tilde{n})}
\label{eq:tildeV}
\end{align}
where
\begin{align}
\tilde{n}
\,=\,
\lceil{
\log_2
{\mathrm{max}}
(r_{\mathrm{in}},r_{\mathrm{out}})
\rceil}
\,.
\end{align}
Here we define the $2^{\tilde{n}} \times 2^{\tilde{n}}$ matrix $\widetilde{V}^{(\tilde{n})}$ with the $r_{\mathrm{out}}\times r_{\mathrm{in}}$ matrix $V$ embedded in the upper left component.
The equation (\ref{eq:tildeV}) implies that we can implement $V^{(n)}$ by the $n$-qubit circuit where the first $(n-\tilde{n})$ qubits are performed by 
measure-and-prepare operations 
independently, and the other qubits are acted on by $\widetilde{V}^{(\tilde{n})}$.
We note that 
$\widetilde{V}^{(\tilde{n})}$ is not unitary in general. 
This means that we need measure-and-prepare operations to implement 
$\widetilde{V}^{(\tilde{n})}$.
We will show an example of implementation in
Section~\ref{sec:numexp}.

\begin{figure*}[t]
\centering
\includegraphics[width=15cm]{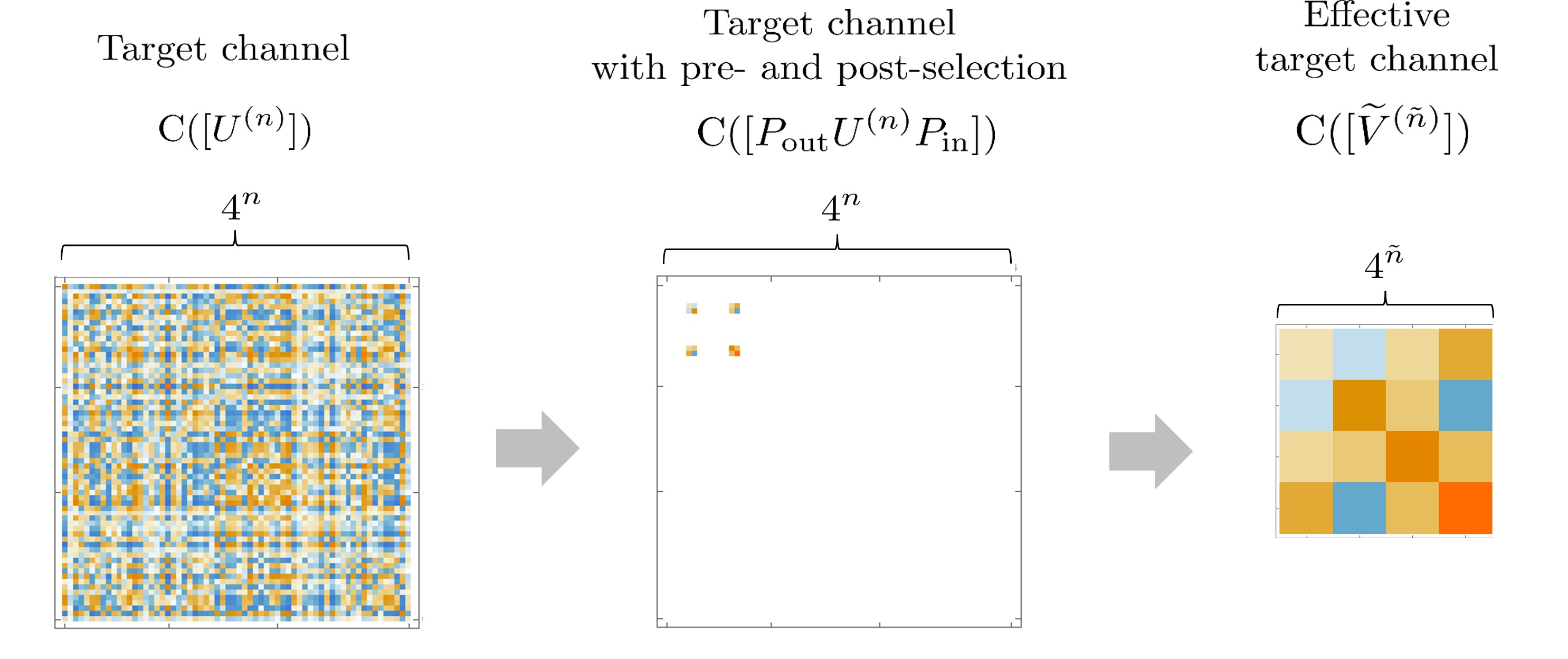}
\caption{Relationship between  
the $n$-qubit channel with and without pre- and post-selection.
We here consider the $n=3$ case where 
$P_{\rm{in}}=\ket{0}\!\!\bra{0}\otimes \ket{0}\!\!\bra{0}\otimes I$ and 
$P_{\rm{out}}=\ket{1}\!\!\bra{1}\otimes \ket{0}\!\!\bra{0}\otimes I$ as an example.
The left, middle and right matrices respectively correspond to the Choi 
matrices of the target unitary without the projection operations $[U^{(n)}]$, 
the target with the projection operators $[P_{\rm{out}}U^{(n)}P_{\rm{in}}]$, 
and the effective target $[\widetilde{V}^{(\tilde{n})}]$.
}
\label{fig:choi}
\end{figure*}

Figure \ref{fig:choi} shows the schematic picture of 
the relationship among $[U^{(n)}]$, 
$[P_{\rm{out}}U^{(n)}P_{\rm{in}}]$,
and $[\widetilde{V}^{(\tilde{n})}]$.
We here consider the $n=3$ case where 
$P_{\rm{in}}=\ket{0}\!\!\bra{0}\otimes \ket{0}\!\!\bra{0}\otimes I$ and 
$P_{\rm{out}}=\ket{1}\!\!\bra{1}\otimes \ket{0}\!\!\bra{0}\otimes I$ as an example.
The left, middle and right matrices in
Fig.~\ref{fig:choi}
correspond to $\mbox{C}([U^{(n)}])$,
$\mbox{C}([P_{\rm{out}}U^{(n)}P_{\rm{in}}])$, and $\mbox{C}([\widetilde{V}^{(\tilde{n})}])$, respectively.
We observe that 
$
\mbox{C}([P_{\rm{out}}U^{(n)}P_{\rm{in}}])$
is the principle submatrix of 
$\mbox{C}([U^{(n)}])$
where the subtraction is specified by $P_{\rm{in}}$ and $P_{\rm{out}}$.
The number of non-trivial components of $\mbox{C}([P_{\rm{out}}U^{(n)}P_{\rm{in}}])$ corresponds to $r_{\rm{out}}\times r_{\rm{in}}$.  
$\mbox{C}([\widetilde{V}^{(\tilde{n})}])$
is constructed by gathering the non-trivial components of $\mbox{C}([P_{\rm{out}}U^{(n)}P_{\rm{in}}])$.
It should be emphasized that 
the matrix size of $\mbox{C}([\widetilde{V}^{(\tilde{n})}])$ is much smaller than $\mbox{C}([{U}^{({n})}])$.

\begin{figure*}[t]
\centering
\includegraphics[width=15cm]{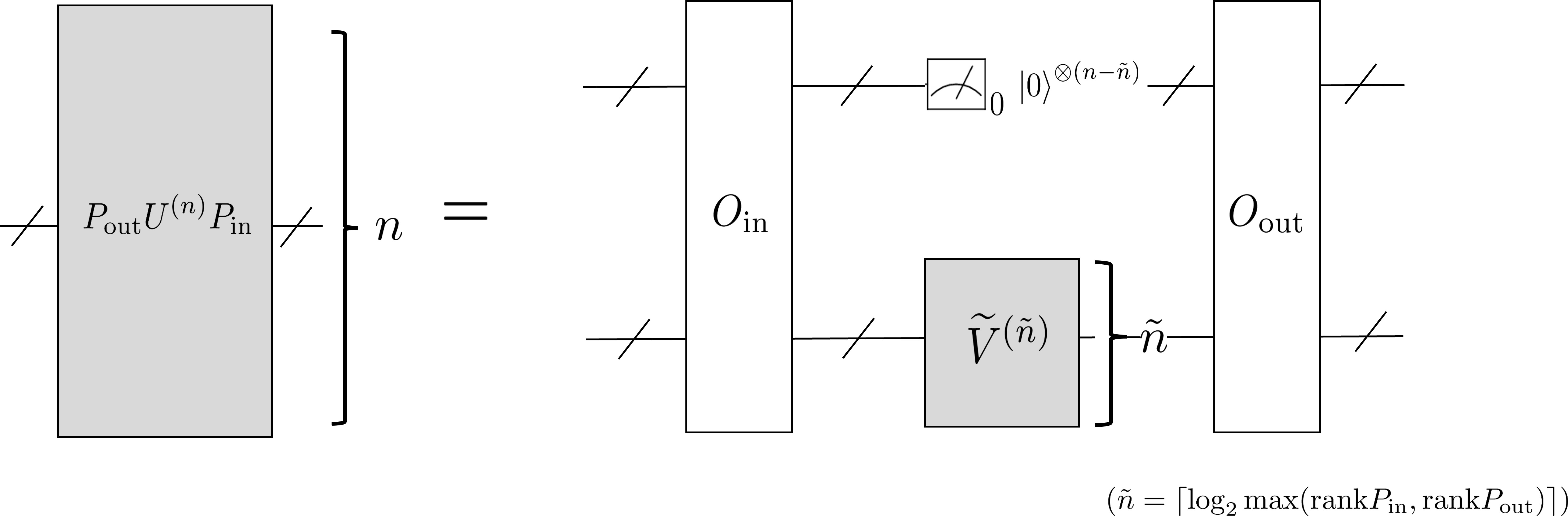}
\caption{
Schematic picture of theorem \ref{thm1}.
The $n$-qubit quantum circuit with the pre- and post-selection (left) can be 
regarded as the quantum circuit shown in the right-handed side. 
$O_{\rm{in}}$ and $O_{\rm{out}}$ 
are the unitary matrices which diagonalize $P_{\rm{in}}$ and $P_{\rm{out}}$ as Eqs.~(\ref{eq:Pin}) and (\ref{eq:Pout}).  
The target unitary dependence is summarized into $\widetilde{V}^{(\tilde{n})}$.
}
\label{fig:thm}
\end{figure*}

Combining Eq.~(\ref{eq:PUP}) and (\ref{eq:tildeV}),
we arrive at the following theorem. 
A schematic picture of this fact is given in Fig.~\ref{fig:thm}.

\begin{thm}
\label{thm1}
Let $P_{\mathrm{in}}$ and $P_{\mathrm{out}}$ be projection matrices with ${\mathrm{rank}}P_{\mathrm{in}}=r_{\mathrm{in}}$ 
and ${\mathrm{rank}}P_{\mathrm{in}}=r_{\mathrm{in}}$. 
Then, the $n$-qubit quantum channel with the pre- and post-selection, 
$[P_{\mathrm{out}}U^{(n)}P_{\mathrm{in}}]$, satisfies
\begin{align}
[P_{\mathrm{out}}U^{(n)}P_{\mathrm{in}}]
\,=\,
[O_{\mathrm{out}}
(\ket{0}\!\!\bra{0}^{\otimes (n-\tilde{n})}\otimes \widetilde{V}^{(\tilde{n})})
O_{\mathrm{in}}]
\,,
\label{eq:thm1}
\end{align}
where $\tilde{n}=\lceil{\log_2{\mathrm{max}}(r_{\mathrm{in}},r_{\mathrm{out}})\rceil}$
and $\widetilde{V}^{(\tilde{n})}$ denotes $2^{\tilde{n}} \times 2^{\tilde{n}}$ matrix.
$O_{\rm{in}}$ and $O_{\rm{out}}$ are known unitary matrices which 
diagonalize $P_{\rm{in}}$ and $P_{\rm{out}}$
as Eqs.~(\ref{eq:Pin}) and (\ref{eq:Pout}). 
$\widetilde{V}^{(\tilde{n})}$ is obtained 
from Eqs.~(\ref{eq:V1}), (\ref{eq:V2}) and (\ref{eq:tildeV}). 
\end{thm}

\subsubsection{Example 1}
\label{sec:exam1}
\begin{figure*}[t]
\centering
\includegraphics[width=15cm]{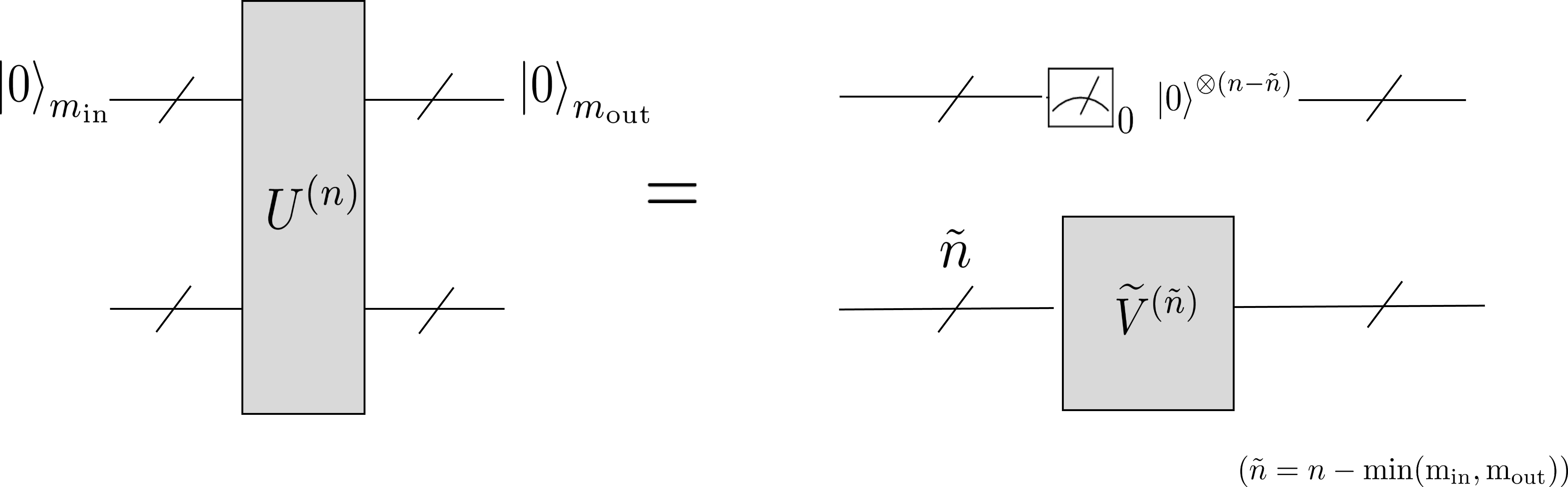}
\caption{
Schematic picture of corollary \ref{cor1}.
This is the specific case of Fig.~\ref{fig:thm} where $P_{\rm{in}}$ and $P_{\rm{out}}$ are given by Eqs.~(\ref{eq:Pin-cor1}) and (\ref{eq:Pout-cor1}), respectively.
}
\label{fig:cor1}
\end{figure*}
Let us consider the case where
the first $m_{\mathrm{in}}$ and $m_{\mathrm{out}}$ qubits of  
the input and output states are specified to be $\ket{0}$ (Figure\,\ref{fig:cor1}).
The pre- and post-selection of the quantum channel is
described by
\begin{align}
P_{\mathrm{in}}
&\,=\,
\ket{0}\!\!\bra{0}^{\otimes m_{\mathrm{in}}}
\otimes I^{\otimes (n-m_{\mathrm{in}})}_{2}
\,=\,
\begin{pmatrix}
    I_{r_{\mathrm{in}}} & \phantom{I_{r_{\mathrm{in}}}}\\
    \phantom{I_{r_{\mathrm{in}}}} & \phantom{I_{r_{\mathrm{in}}}}\\
\end{pmatrix}
\,,
\label{eq:Pin-cor1}\\
P_{\mathrm{out}}
&\,=\,
\ket{0}\!\!\bra{0}^{\otimes m_{\mathrm{out}}}
\otimes I^{\otimes (n-m_{\mathrm{out}})}_{2}
\,=\,
\begin{pmatrix}
    I_{r_{\mathrm{out}}} & \phantom{I_{r_{\mathrm{out}}}}\\
    \phantom{I_{r_{\mathrm{out}}}} & \phantom{I_{r_{\mathrm{out}}}}\\
\end{pmatrix}
\,,
\label{eq:Pout-cor1}
\end{align}
with
\begin{align}
r_{\mathrm{in}}
\,=\,
2^{n-m_{\mathrm{in}}}
\,,\quad
r_{\mathrm{out}}
\,=\,
2^{n-m_{\mathrm{out}}}
\,.
\end{align}

Since both the $P_{\mathrm{in}}$ and $P_{\mathrm{out}}$ have the same 
structure as $P_{d,\mathrm{in}}$ and $P_{d,\mathrm{out}}$ in
Eqs.~(\ref{eq:Pin}) and (\ref{eq:Pout}), we can take
\begin{align}
O_{\mathrm{in}}
=
O_{\mathrm{out}}
=
I_{N}
\,.
\end{align}

The explicit form of $\widetilde{V}^{(\tilde{n})}$ depends on whether 
$m_{\mathrm{out}}$ is larger than $m_{\mathrm{in}}$ or not. 
If $m_{\mathrm{out}}\leq m_{\mathrm{in}}$, then 
$\tilde{n}=n-m_{\mathrm{out}}$ and 
\begin{align}
\widetilde{V}^{(\tilde{n})}
\,=\,
\begin{pmatrix}
    (v_1)_{1} & \cdots & (v_{r_{\mathrm{in}}})_{1} & 0 & \cdots & 0 \\
    \vdots &        & \vdots &  \vdots &        & \vdots \\
           & \ddots & \vdots &         & \ddots &        \\
    \vdots &        & \vdots &  \vdots &        & \vdots \\
    (v_1)_{r_{\mathrm{out}}} & \cdots & (v_{r_{\mathrm{in}}})_{r_{\mathrm{out}}} & 0 & \cdots & 0 \\
\end{pmatrix}
\,.
\label{eq:V-exam1-1}
\end{align}
If $m_{\mathrm{out}}> m_{\mathrm{in}}$, then 
$\tilde{n}=n-m_{\mathrm{in}}$ and 
\begin{align}
\widetilde{V}^{(\tilde{n})}
\,=\,
\begin{pmatrix}
  (v_1)_1 & \cdots  &   & \cdots & (v_{r_{\mathrm{in}}})_1 \\
  \vdots &   &  \ddots &  &  \vdots \\
  (v_1)_{r_{\mathrm{out}}} & \cdots & & \cdots & (v_{r_{\mathrm{in}}})_{r_{\mathrm{out}}}  \\
  0 & \cdots &  & \cdots & 0 \\
  \vdots &  & \ddots &  & \vdots \\
  0 & \cdots &  & \cdots & 0 \\
\end{pmatrix}
\,.
\label{eq:V-exam1-2}
\end{align}
In both cases, the matrix element is given as
\begin{align}
(v_i)_{j}
=
\sbraket{E_{ji}}{U^{(n)}}
\,.
\label{eq:v-exam1}
\end{align}

We then arrive at the following corollary:
\begin{cor}
\label{cor1}
The $n$-qubit quantum channel $[P_{\mathrm{out}}U^{(n)}P_{\mathrm{in}}]$ where $P_{\mathrm{in}}$ 
and $P_{\mathrm{out}}$ are defined by Eqs.~(\ref{eq:Pin-cor1}) and (\ref{eq:Pout-cor1}) satisfies
\begin{align}
[P_{\mathrm{out}}U^{(n)}P_{\mathrm{in}}]
\,=\,
[\ket{0}\!\!\bra{0}^{\otimes(n-\tilde{n})}
\widetilde{V}^{(\tilde{n})}]
\end{align}
where $\tilde{n}=n-\min(m_{\mathrm{in}},m_{\mathrm{out}})$.
If $m_{\mathrm{out}}\leq m_{\mathrm{in}}$,
$\tilde{V}^{(\tilde{n})}$ is calculated from Eqs.~(\ref{eq:V-exam1-1}) and (\ref{eq:v-exam1}). If $m_{\mathrm{out}}>m_{\mathrm{in}}$, it is calculated from
Eqs.~(\ref{eq:V-exam1-2}) and (\ref{eq:v-exam1}). 
\end{cor}
The schematic picture of Corollary \ref{cor1} is shown in Fig.~\ref{fig:cor1}.

\subsubsection{Example 2}
\label{sec:exam-hhl}

\begin{figure*}[t]
\centering
\includegraphics[width=15cm]{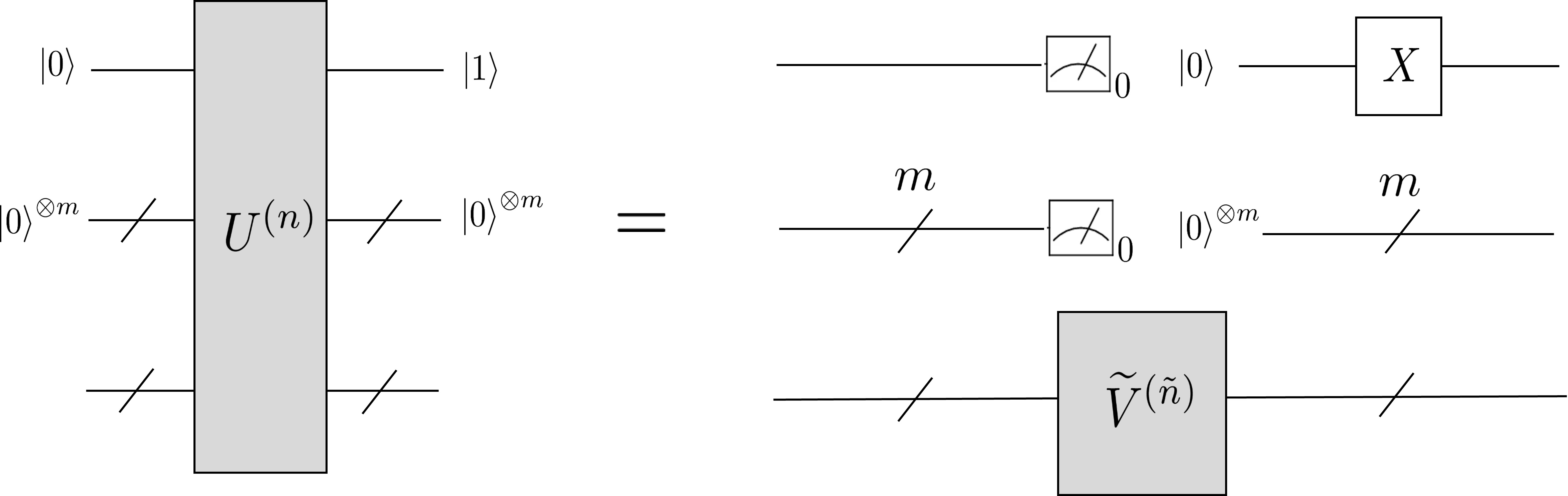}
\caption{
Schematic picture of Corollary \ref{cor2}.
This is the specific case of Fig.~\ref{fig:thm} where $P_{\rm{in}}$ and $P_{\rm{out}}$ are given by Eqs.~(\ref{eq:Pin-cor2}) and (\ref{eq:Pout-cor2}), respectively.
HHL circuit (which is discussed in Section~\ref{sec:numexp}) falls into this case.
}
\label{fig:cor2}
\end{figure*}

We next consider the case depicted in Fig.~\ref{fig:cor2}: 
The first $(m+1)$-qubits of the input state are set to be $\ket{0}$. 
For the output state, the first qubit is specified as $\ket{1}$ and the second 
to $(m+1)$th qubit are specified as $\ket{0}$. 
The quantum circuit of the HHL algorithm, which is discussed later, falls into 
this type.

The pre- and post-selection of the quantum channel is described by
\begin{align}
P_{\mathrm{in}}
&=
\ket{0}\!\!\bra{0}^{\otimes (m+1)}
\otimes 
I^{\otimes(n-m-1)}
\,=\,
\begin{pmatrix}
    I_{r_{\mathrm{in}}}           & \phantom{I_{r_{\mathrm{in}}}}\\
    \phantom{I_{r_{\mathrm{in}}}} & \phantom{I_{r_{\mathrm{in}}}}\\
\end{pmatrix}
\,,
\label{eq:Pin-cor2}\\
P_{\mathrm{out}}
&=
\ket{1}\!\!\bra{1}
\otimes 
\ket{0}\!\!\bra{0}^{\otimes m}
\otimes 
I^{\otimes(n-m-1)}
\nonumber\\
&\,=\,
\begin{pmatrix}
0_{2^{n-1}}               & \phantom{I_{r_{\mathrm{in}}}} & \phantom{I_{r_{\mathrm{in}}}} \\
\phantom{I_{r_{\mathrm{in}}}} & I_{r_{\mathrm{out}}}          & \phantom{I_{r_{\mathrm{in}}}} \\
\phantom{I_{r_{\mathrm{in}}}} & \phantom{I_{r_{\mathrm{in}}}} & \phantom{I_{r_{\mathrm{in}}}}
\end{pmatrix}
\,,
\label{eq:Pout-cor2}
\end{align}
with $0_m$ being $m\times m$ zero matrix and
\begin{align}
r_{\mathrm{in}}
\,=\,
r_{\mathrm{out}}
\,=\,
2^{n-m-1}
\,:=\,
r\,.
\end{align}

Since $P_{\mathrm{in}}$ is the matrix with the same structure as $P_{d,\mathrm{in}}$ (Eq.~(\ref{eq:Pin})), we take
\begin{align}
O_{\mathrm{in}}
=
I_N\,.
\end{align}
On the other hand, 
$P_{\mathrm{out}}$ is the diagonal matrix but the structure is different from $P_{d,\mathrm{out}}$ (Eq.~(\ref{eq:Pout})).
To make $P_{\mathrm{out}}$ have the same structure as $P_{d,\mathrm{out}}$, 
we perform unitary transformation by
\begin{align}
O_{\mathrm{out}}
&\,=\,
X\otimes I^{\otimes(n-1)}
\,.
\end{align}

The effective number of qubits $\tilde{n}$ is given as
\begin{align}
\tilde{n}
=
n-m-1
\,.
\end{align}
The effective operation  $\widetilde{V}^{(\tilde{n})}$ is calculated as
\begin{align}
\widetilde{V}^{(\tilde{n})}
\,=\,
\begin{pmatrix}
(v_1)_1 & \cdots & (v_1)_{r} \\ 
\vdots & \ddots & \vdots \\
(v_{r})_1 & \cdots & (v_{r})_{r} \\ 
\end{pmatrix}
\,,
\label{eq:V-cor2}
\end{align}
where
\begin{align}
(v_i)_{j}
&\,=\,
\sbraket{E_{ji}}{O_{\mathrm{out}}U^{(n)}}
\nonumber\\
&\,=\,
\sbraket{O^\dagger_{\mathrm{out}}E_{ji}}{U^{(n)}}
\nonumber\\
&\,=\,
\sbraket{E_{k_j i}}{U^{(n)}}
\,,
\label{eq:v-cor2}
\end{align}
with $k_j = (2^{n-1} + j - 1)\, \mathrm{mod}\, 2^n + 1$.

We arrive at the following corollary:
\begin{cor}
\label{cor2}
The $n$-qubit quantum channel $[P_{\mathrm{out}}U^{(n)}P_{\mathrm{in}}]$ 
where $P_{\mathrm{in}}$ and $P_{\mathrm{out}}$ are defined 
by Eqs.~(\ref{eq:Pin-cor2}) and (\ref{eq:Pout-cor2}) can be equivalently expressed as
\begin{align}
&[P_{\mathrm{out}}U^{(n)}P_{\mathrm{in}}]
\nonumber\\
&\,=\,
[
(X\otimes I^{\otimes(n-1)})
(\ket{0}\!\!\bra{0}^{\otimes(n-\tilde{n})}
\widetilde{V}^{(\tilde{n})})
]\,,
\end{align}
where $\tilde{n}=n-m$ and
$\widetilde{V}^{(\tilde{n})}$
is given by 
Eqs.~(\ref{eq:V-cor2}) and (\ref{eq:v-cor2})
\end{cor}
The schematic picture of corollary \ref{cor2} is 
shown in Fig.~\ref{fig:cor2}.

\subsection{Gate decomposition with pre- and post-selection}

As we discussed above, the $n$-qubit channel with pre- and post-selection, 
$[P_{\mathrm{out}}U^{(n)}P_{\mathrm{in}}]$, is characterized by 
$\tilde{n}$-qubit channel, $[\widetilde{V}^{(\tilde{n})}]$.
We here perform the gate decomposition (\ref{eq:gate-cutting}) 
to $[\widetilde{V}^{(\tilde{n})}]$
as
\begin{align}
[\widetilde{V}^{(\tilde{n})}]
\,=\,
\sum_{i=1}^{\tilde{D}}
c_i\,
[
V^{(1)}_{i_1}\otimes
\cdots\otimes
V^{(1)}_{i_{\tilde{n}}}
]\,,
\label{eq:tildeV-decomp}
\end{align}
with 
$i=\sum_{k=1}^{\tilde{n}}(\prod_{k'=1}^k D_{k'-1})i_k$.
We note that 
$\widetilde{D}$ needs to be only larger than $16^{\tilde{n}}$
which is generally smaller than $N^4=16^n$.

Combining Eq.~(\ref{eq:tildeV-decomp}) with Theorem \ref{thm1}, we find 
the following fact:

\begin{thm}
Let $P_{\mathrm{in}}$ and $P_{\mathrm{out}}$ be projection matrices with ${\mathrm{rank}}P_{\mathrm{in}}=r_{\mathrm{in}}$ 
and ${\mathrm{rank}}P_{\mathrm{out}}=r_{\mathrm{out}}$.
The $n$-qubit quantum channel with the pre- and post-selection, 
$[P_{\mathrm{out}}U^{(n)}P_{\mathrm{in}}]$, is decomposed as
\begin{align}
&[P_{\mathrm{out}}U^{(n)}P_{\mathrm{in}}]
\nonumber\\
&\,=\,
\sum_{i=1}^{\widetilde{D}}
c_i\,
[
O_{\mathrm{out}}
(
\ket{0}\!\!\bra{0}^{\otimes(n-\tilde{n})}
\otimes
V^{(1)}_{i_1}\otimes
\cdots\otimes
V^{(1)}_{i_{\tilde{n}}}
)
O_{\mathrm{in}}
]\,,
\label{eq:thm2}
\end{align}
where 
$i=\sum_{k=1}^{\tilde{n}}(\prod_{k'=1}^k D_{k'-1})i_k$ with $D_0=1$,
$\tilde{n}=\lceil{
\log_2{\mathrm{max}}(r_{\mathrm{in}},r_{\mathrm{out}})
\rceil}$, 
and $\{V^{(1)}_{i_a}\,,(a=1,\cdots,\tilde{n})\}$ denotes
the set of linearly independent one-qubit quantum channels.
We need $\widetilde{D}\geq 16^{\tilde{n}}$ to satisfy the equality. 
$O_{\mathrm{in}}$ and $O_{\mathrm{out}}$ are the unitary matrices defined as
Eqs.~(\ref{eq:Pin}) and (\ref{eq:Pout}).
\end{thm}

As the demonstration, 
we perform the gate decomposition (\ref{eq:thm2}) of 
Toffoli and the three-qubit QFT gate with
pre- and post-selection. 
Here we consider the case where $P_{\rm{in}}=\ket{0}\!\!\bra{0}^{\otimes m_{\rm{in}}}
\otimes I^{(3-m_{\rm{in}})}$ and 
$P_{\rm{out}}=\ket{0}\!\!\bra{0}^{\otimes m_{\rm{out}}}
\otimes I^{(3-m_{\rm{out}})}$
with $0\leq m_{\rm{in}}, m_{\rm{out}} \leq 3$.
This is the case discussed in Section~\ref{sec:exam1}. 
To obtain the coefficients $\{c_i\}$, 
we calculate $\widetilde{V}^{(\tilde{n})}$ using Corollary~\ref{cor1} 
and solve Eq.~(\ref{eq:tildeV-decomp}){\footnote{
The detail on how we solve Eq.~(\ref{eq:tildeV-decomp}) is as follows. 
We first solve $
{\mbox{C}}
([\widetilde{V}^{(n)}])
=
\sum_{i=1}^{\widetilde{D}}
c'_i 
{\mbox{C}}(
[
V'^{(1)}_{i_1}\otimes
\cdots\otimes
V'^{(1)}_{i_{\tilde{n}}}
]
)
$ with $\widetilde{D}=16^{\tilde{n}}$
and
$\{{V}'^{(1)}_{i_a}\}$ being
the set of one-qubit operations given in Table \ref{tab:basis}.
The effective target matrix
$\widetilde{V}^{(n)}$ is calculated by Eqs.~(\ref{eq:V-exam1-1}) and (\ref{eq:V-exam1-2}).
We find that the sum of the coefficients $\{c'_i\}$ is positive and real but deviates from unity, $\sum_{i=1}c'_i=c'\neq 1$.
To obtain the normalized coefficients $\{c_i\}$, 
we define $c_i=c'_i/c'$ so that $\sum_i c_i=1$. 
We also define ${V}^{(1)}_{i_a}=\sqrt{c}'{V}'^{(1)}_{i_a}$.
We then obtain 
$\mbox{C}([\widetilde{V}^{(n)}])
=
\sum_{i=1}^{\widetilde{D}}
c_i
\mbox{C}(
[
V^{(1)}_{i_1}\otimes
\cdots\otimes
V^{(1)}_{i_{\tilde{n}}}
]
=
\mbox{C}(
\sum_{i=1}^{\widetilde{D}}
c_i
[
V^{(1)}_{i_1}\otimes
\cdots\otimes
V^{(1)}_{i_{\tilde{n}}}
]
)$ 
where we use the affine property of 
Jamiolkowski–Choi isomorphism in the second equality.
We then arrive at Eq.~(\ref{eq:tildeV-decomp}).
}}.
The resulting sampling overheads $\gamma=\sum_i|c_i|$ are presented in 
Tables~\ref{tab:Toffoli} and \ref{tab:QFT}.
\begin{table}
\caption{
Sampling overhead $\gamma=\sum_{i=1}|c_i|$ for 
the decomposition of Toffoli gate 
with pre- and post-selection.
$\ket{abc}=\ket{a}\otimes
\ket{b}\otimes
\ket{c}$ and 
$\ket{*}$ denotes an arbitrary 
one-qubit pure state.
}
\label{tab:Toffoli}
\begin{tabular}{c|cccc}
\diagbox{in}{out}& 
$\ket{***}$ & $\ket{0**}$ & $\ket{00*}$ & $\ket{000}$\\
\hline
$\ket{***}$ & 37 & 37 & 37 & 37\\
$\ket{0**}$ & 37 & 1 & 1 & 1\\
$\ket{00*}$ & 37 & 1 & 1 & 1 \\
$\ket{000}$ & 37 & 1 & 1 & 1 \\
 \end{tabular}
\end{table}
\begin{table}
\caption{
Sampling overhead $\gamma=\sum_{i=1}|c_i|$ for 
the decomposition of the three-qubit QFT gate 
with pre- and post-selection. $\ket{abc}=\ket{a}\otimes
\ket{b}\otimes
\ket{c}$ and 
$\ket{*}$ denotes an arbitrary 
one-qubit pure state.}
\label{tab:QFT}
\begin{tabular}{c|cccc}
\diagbox{in}{out}& 
$\ket{***}$ & $\ket{0**}$ & $\ket{00*}$ & $\ket{000}$\\
\hline
$\ket{***}$ & 261.43 & 261.43 & 261.43  & 261.43\\
$\ket{0**}$ & 261.43 & 261.43  & 16.63  & 16.63\\
$\ket{00*}$ & 261.43 & 16.63   & 1.64   & 1.64 \\
$\ket{000}$ & 261.43 & 16.63 & 1.64 & 1 \\
\end{tabular}
\end{table}
We observe that 
the pre- and post-selection reduces 
the cost of the gate decomposition.

\section{Numerical experiments}
\label{sec:numexp}

The gate decomposition method allows us to simulate a complicated quantum 
circuit by sampling the outputs of relatively simpler quantum circuits. 
The gate decomposition is therefore expected to reduce the impact of errors 
during the quantum computation. 
In this section, we actually demonstrate this benefit of the gate decomposition.

We here study the HHL algorithm \cite{harrow2009quantum, ambainis2010variable,childs2017quantum, liu2023towards, dervovic2018quantum, morrell2021step, tosti2022review}. 
Since the HHL algorithm requires the pre- and post-selections on many ancilla 
qubits, the advantage of our decomposition method should be manifest{\footnote{
There may exist quantum algorithms which can be more efficiently decomposed by our method. 
We will discuss such candidates in Section \ref{sec:conculusion}.
}}.

\subsection{Harrow-Hassidim-Lloyd (HHL) algorithm}

Here we briefly describe the HHL algorithm.
This is a quantum algorithm to solve the linear equation, $A\vec{x}=\vec{b}$. 
Here $A$ and $\vec{b}$ denote a $\tilde{N}\times \tilde{N}$ complex matrix and a 
$\tilde{N}$-dimensional complex vector, respectively. 
The algorithm produces an approximating solution of a certain type of function 
of $\vec{x}$ (e.g., $\vec{x}^\dagger H \vec{x}$ with $H$ a Hermitian matrix), 
with running time of $\mathcal{O}({\rm{poly}}{\rm{log}}{\tilde{N}})$ when $A$ 
is sparse, while any classical computation needs $\mathcal{O}({\tilde{N}})$ 
running time.

\begin{figure*}[t]
\centering
\includegraphics[width=12cm]{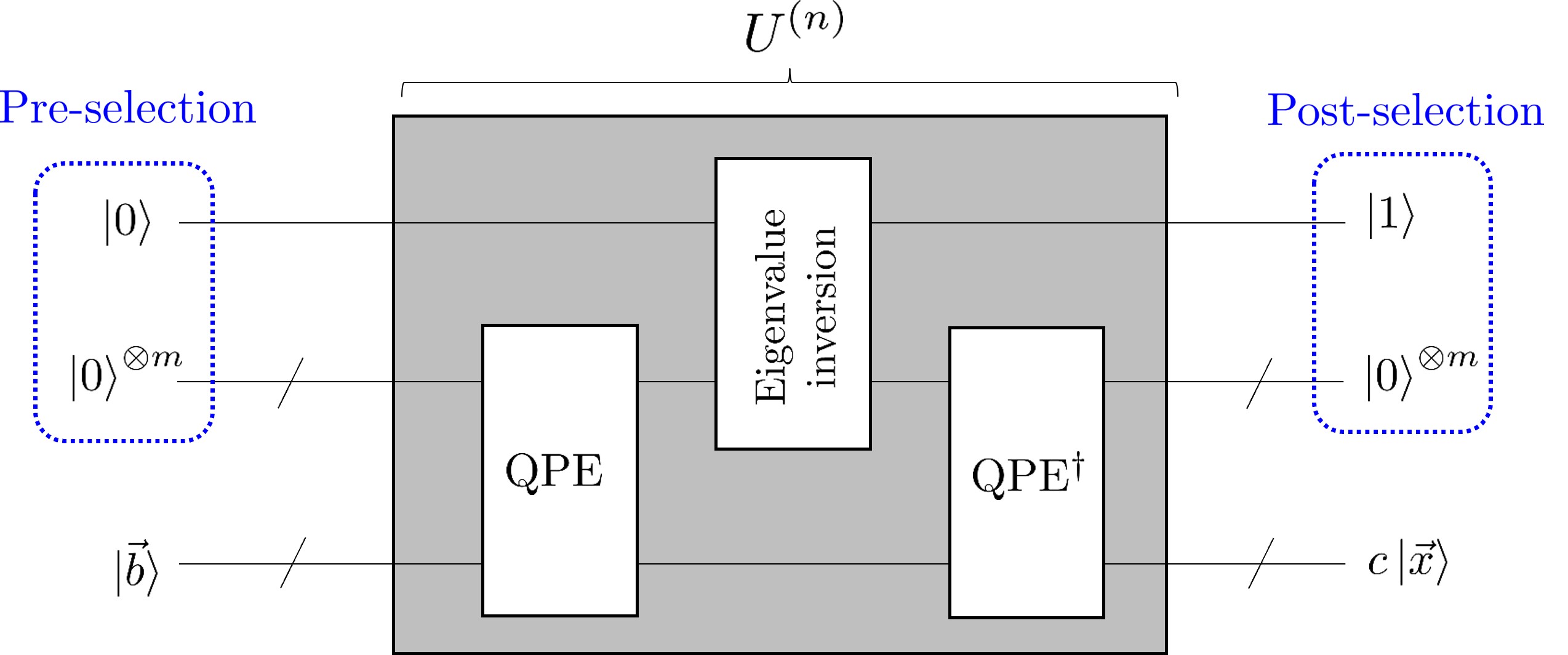}
\caption{
Quantum circuit of the HHL algorithm for solving  $A\vec{x}=\vec{b}$. 
The first $(m+1)$ qubits are used as ancilla qubits which are initialized to 
be $\ket{0}^{\otimes(m+1)}$ (pre-selection).
The unitary operation $U^{(n)}$ has three steps: 
(i) Quantum Phase Estimation (QPE), 
(ii) conditional rotation to encode the inverse of the eigenvalues of $A$,
and (iii) uncomputation (QPE$^\dag$).
With this post-selection of the ancilla qubits as the 
$\ket{1}\otimes\ket{0}^{\otimes m}$, we obtain the solution state $\vec{x}$
up to a normalization factor $c$ in the working qubits.
}
\label{fig:hhl-circ}
\end{figure*}

The basic quantum circuit of the HHL is shown in Fig.~\ref{fig:hhl-circ}.
The algorithm needs $n=\tilde{n}+m+1$ qubits where $\tilde{n}$-qubits are used 
for the amplitude encoding of $\vec{b}$ and the $(m+1)$-qubits are used as ancilla. 
The $m$-ancilla qubits are used to store the binary representation of the 
eigenvalues of $A$, and the rest one ancilla qubit is used to encode the inverse 
of the eigenvalues of the matrix. 
All ancilla qubits are set to be $\ket{0}$ at the beginning (pre-selection),
and they are post-selected as $\ket{1}\otimes\ket{0}^{\otimes m}$. 
With this post-selection, we obtain the solution state $\ket{\vec{x}}$
up to a normalization factor $c$ in the working qubits.

In our terminology, the quantum channel of HHL corresponds to the case 
discussed in Section~\ref{sec:exam-hhl}. 
The target unitary operation $U^{(n)}$ consists of the following three steps.
First, it applies Quantum Phase Estimation (QPE) to obtain the binary 
representation of the eigenvalues of $A$. 
The eigenvalues are stored in the middle $m$-ancilla qubits.
Second, the conditional rotation is performed to get the inverse eigenvalues 
of $A$. 
The information of the inverse eigenvalues is stored in the coefficient of 
$\ket{1}$ of the first ancilla qubit. 
Third, we apply QPE$^\dag$ for the uncomputation.

It should be noted that the HHL requires huge resources in general. 
As we will see below, even in the simplest $(\tilde{N}=2)$ case, the quantum 
circuit typically has hundreds of depths and requires tens of CNOTs. 
This implies that the quantum computation seriously suffers from huge errors. 
In the next subsection, we perform the gate decomposition of the HHL circuit 
and demonstrate that it drastically reduces the effect of error.

\subsection{Gate decomposition of HHL circuit}

Here we study the HHL algorithm for solving $A\vec{x}=\vec{b}$ with
\begin{align}
    A \,=\,
    \begin{pmatrix}
     1 & -1/3 \\
     -1/3 & 1
    \end{pmatrix}
    \,,
    \quad
    \vec{b}
    =
    \begin{pmatrix}
    1 \\ 
    0
    \end{pmatrix}
    \,.
\end{align}
This corresponds to the case of $\tilde{n}=1$ ($\tilde{N}=2$).

We used Qiskit~\cite{Qiskit} to simulate the HHL circuit. 
In particular, we generated the quantum circuit by using {\tt{linear\_solvers}} 
package in Qiskit. 
The depth of the resulting circuit is 212, and it contains 108 CNOT gates. 
Also, $m=3$ ancilla qubits are used to store the eigenvalues of $A$. 
The quantum circuit has, therefore, 
$n=1+3+1=5$ qubit in total\footnote{
As noted in the tutorial, the quantum circuit generated with Qiskit is not the 
optimal one. 
However, we simply used the naive circuit obtained from {\tt{linear\_solvers}} 
package for the proof-of-concept demonstration. 
}.

First, we show the simulation result of the HHL algorithm without the gate 
decomposition. 
We estimated the output state by performing the state tomography simulation (with 
10000 shots) using Qiskit. 
The post-selection is implemented by extracting the relevant components of the 
output state. 
If the algorithm correctly works, the extracted density matrix $\rho_{\rm{out}}$ 
ends up with $\rho_{\rm{sol}}=\ket{\vec{x}}\!\!\bra{\vec{x}}$ with $\ket{\vec{x}}=\ket{A^{-1}\vec{b}}=[9/8,3/8]^T$ up to constant. 
To evaluate the performance of the algorithm, we used {\tt{state\_fidelity}} 
function in Qiskit to compute the fidelity $\mathcal{F}$ between the normalized 
output state $\hat{\rho}_{\rm{out}}={\rho}_{\rm{out}}/\mbox{Tr}({\rho}_{\rm{out}})$ 
and the solution state $\hat{\rho}_{\rm{sol}}={\rho}_{\rm{sol}}/\mbox{Tr}({\rho}_{\rm{sol}})$. 
Also, we consider the case with and without quantum noise. 
In the simulation with the noise, we assume that the one-qubit depolarization 
error occurs with probability $p_{\rm{local}}=0.001$ for one-qubit gates 
and $p_{\rm{CNOT}}=0.005,~0.01$ for CNOT gates. 
The results of the fidelity $\mathcal{F}$ are shown in the top row of 
Table~\ref{tab:cut-hhl} indicated by "w/o decomposition". 
In the case without noise (i.e., $p_{\rm{local}}=p_{\rm{CNOT}}=0$), 
we observe $\mathcal{F}\simeq 0.99$, 
which ensures that the HHL algorithm works well; note that the small deviation 
from the ideal value $\mathcal{F}=1$ is due to the statistical error in the 
state tomography. 
In contrast, the fidelity gets worse as $\mathcal{F}\simeq 0.86$ and 
$\mathcal{F}\simeq0.75$ for the cases $p_{\rm{CNOT}}=0.005$ and $p_{\rm{CNOT}}=0.01$, respectively. 
This is simply because there are too many CNOT gates in the HHL circuit 
and they all seriously suffer from the noise. 

\begin{table*}
\caption{
Fidelity $\mathcal{F}$ between the output state of the HHL and the solution state 
with and without the gate decomposition. 
If the HHL algorithm perfectly works, the value of fidelity becomes unity. 
The noise is assumed to be the one-qubit depolarization error with the error 
probability $p_{\rm{local}}=0.001$ for one-qubit gates 
and $p_{\rm{CNOT}}=0.005$ and $0.01$ for CNOT gate. 
The noise-free case $p_{\rm{local}}=p_{\rm{CNOT}}=0$ is also shown. 
}
\label{tab:cut-hhl}
\begin{tabular}{c||c|c|c}
 &  ~~Noise-free~~ & ~~$p_{\rm{CNOT}}=0.005$~~ & ~~$p_{\rm{CNOT}}=0.01$~~ \\
\hline
\hline
 ~~w/o decomposition~~ & 0.99 &  0.86 &  0.75\\
 ~~w/ decomposition~~ & 0.99 &  0.99 & 0.99\\
\end{tabular}
\end{table*}

We next study the performance of the gate-decomposed HHL circuit. 
As mentioned before, the quantum channel of the HHL algorithm is included in the 
class of channels discussed in Section~\ref{sec:exam-hhl}. 
Specifically, $\tilde{n}=1$ in our case. 
We first compute the effective operation $\widetilde{V}^{(\tilde{n})}$ given in 
Eqs.~(\ref{eq:V-cor2}) and (\ref{eq:v-cor2}). 
Then, Eq.~(\ref{eq:tildeV-decomp}) is solved to obtain the coefficients $\{c_i\}$, 
where the basis in Table \ref{tab:basis} is employed. 
The result is that 
\begin{align}
[\widetilde{V}^{(\tilde{n})}]
\,=\,
\frac{3}{8}[I]
-\frac{1}{8}[X]
+\frac{6}{8}[\pi_X]
\,,
\label{eq:decomp-hhl-tildeV}
\end{align}
which leads to
\begin{align}
[&P_{\rm{out}}U^{(n)} P_{\rm{in}}]
\nonumber\\
\,=
&\,
\frac{3}{8}
[
(X\otimes I\otimes I \otimes I\otimes I)
(\pi_Z\otimes \pi_Z\otimes\pi_Z\otimes\pi_Z\otimes I)
]
\nonumber\\
&\,
-\frac{1}{8}
[
(X\otimes I\otimes I \otimes I\otimes I)
(\pi_Z\otimes \pi_Z\otimes\pi_Z\otimes \pi_Z\otimes X)
]
\nonumber\\
&\,
+\frac{6}{8}
[
(X\otimes I\otimes I \otimes I\otimes I)
(\pi_Z\otimes \pi_Z\otimes\pi_Z\otimes \pi_Z\otimes \pi_X)
]
\,.
\label{eq:decomp-hhl}
\end{align}
Therefore, notably, the original HHL circuit, which is of depth 212 and contains 
108 CNOT gates, can be decomposed to only three channels each of which is of 
depth 2 and does not contains any CNOT gate. 
Because those three channels are expected to be almost free from noise, nearly 
ideal simulation might be possible. 
Furthermore, fortunately, the number of decomposition is only $\tilde{D}=3$, while 
in general $\tilde{D}\geq 16^{\tilde{n}}=16$ is required for the case $\tilde{n}=1$ 
(note thus that 13 of total 16 coefficients $\{c_i\}$ are all zero). 
As a result, the sampling overhead is only $\gamma=5/4$, meaning that the virtual 
simulation via the channel decomposition can be efficiently executed with low 
sampling overhead.

We also calculate the sampling costs for particular larger-dimensional cases 
where 
\begin{align}
    A &\,=\,
    \begin{pmatrix}
     1 & -\frac{1}{3} & 0 & 0 \\
     -\frac{1}{3} & 1 & -\frac{1}{3} & 0 \\
     0 & -\frac{1}{3} & 1 & -\frac{1}{3}  \\
     0 & 0 & -\frac{1}{3} & 1 
    \end{pmatrix}
    \,,
    \label{eq:A2}\\
    A &\,=\,
    \begin{pmatrix}
     1 & -\frac{1}{3} & 0 & 0 & 0 & 0 & 0 & 0 \\
     -\frac{1}{3} & 1 & -\frac{1}{3} & 0 & 0 & 0 & 0 & 0 \\
     0 & -\frac{1}{3} & 1 & -\frac{1}{3} & 0 & 0 & 0 & 0 \\
     0 & 0 & -\frac{1}{3} & 1 & -\frac{1}{3} & 0 & 0 & 0 \\
     0 & 0 & 0 & -\frac{1}{3} & 1 & -\frac{1}{3} & 0 & 0 \\
     0 & 0 & 0 & 0 & -\frac{1}{3} & 1 & -\frac{1}{3} & 0 \\
     0 & 0 & 0 & 0 & 0 & -\frac{1}{3} & 1 & -\frac{1}{3} \\
     0 & 0 & 0 & 0 & 0 & 0 & -\frac{1}{3} & 1 \\
    \end{pmatrix}
    \,,
    \label{eq:A3}
\end{align}
and $\vec{b}=(1,0,\cdots,0)^T$. 
These cases respectively correspond to $\tilde{n}=2$ and
$\tilde{n}=3$.
The resulting sampling costs are 
$\gamma\simeq 4.7$ for Eq.~(\ref{eq:A2}) and 
$\gamma\simeq 16.5$ for Eq.~(\ref{eq:A3}).
Unfortunately, we cannot obtain the analytic form of the decomposition and accordingly the $\gamma$ factors in 
$\tilde{n}\geq 2$ cases. 
This numerical results implies that the sampling cost 
scales exponentially with $\tilde{n}$, 
but the scaling is much milder than $16^{n}$.

The decomposing channels represented in the right-hand side of 
Eq.~(\ref{eq:decomp-hhl}) are products of one-qubit channels and thus easy 
to implement. 
But implementation of the developed channel is easier than expected, 
because the ancilla qubits are initialized as $\ket{0000}$
in the HHL algorithm.
Actually, in this case, the operation on the ancilla qubits is trivial and thus 
can be ignored, meaning that we only need to implement the one-qubit decomposed 
gate (\ref{eq:decomp-hhl-tildeV}). 
$[\pi_X]$ can be implemented as measure-and-prepare channel because $\pi_X=\ket{+}\!\!\bra{+}$.

We can now give a quantitative discussion on the benefit of the gate decomposition, 
from the view of noise-tolerance property. 
Let us assume the same error model as in the analysis without the gate decomposition. 
We summarize the state fidelity with the gate decomposition in the bottom row of 
Table~\ref{tab:cut-hhl}. 
Likewise the w/o decomposition case, $\mathcal{F}\simeq 0.99$ is achieved in the 
noise-free case $p_{\rm{local}}=p_{\rm{CNOT}}=0$, implying that the gate decomposition is successfully executed. 
A notable difference to the previous case is found in the noisy cases $p_{\rm{local}}=0.001$, $p_{\rm{CNOT}}=0.005$ and 
$p_{\rm{CNOT}}=0.01$. 
That is, even in this case the state fidelity $\mathcal{F}$ remains close 
to unity, demonstrating the effectiveness of the gate decomposition method against 
noise.

\section{Conclusion}
\label{sec:conculusion}

The channel-decomposition technique is useful, 
but it suffers from the exponential increase of the number of decompositions, or equivalently the sampling overhead $\gamma$, 
with respect to the size of the target channel. 
This paper proposed a general method for reducing this cost by focusing on 
channels with pre- and post-selection, i.e., the channels whose input and output 
are restricted to fixed set of states. 
We proved that the overhead is determined by the rank of projection operators specifying 
the input and output states. 
Also, the form of decomposition under the pre- and post-selection is explicitly 
given. 
As a demonstration, we applied the method to decompose the unitary operator for 
the HHL algorithm, showing an efficient channel decomposition (or equivalently low 
sampling overhead) thanks to the pre- and post-selection.

There are many examples of channels with pre- and post-selection to which the 
proposed method is applicable. 
A typical example is a gate for probabilistic state preparation, because the quantum operation is embedded into the larger
unitary gate and is implemented by pre- and post-selections 
of ancilla qubits.
The probabilistic imaginary time evolution, which can be used for quantum chemistry, 
also falls into this class \cite{kosugi2022imaginary, Liu2021-vf, Lin2021-er}. 
The differential equation solvers 
\cite{berry2014high,berry2017quantum,xin2020quantum,liu2021efficient,an2022theory}, 
which can be used for finance \cite{rebentrost2018quantum, miyamoto2021pricing}, 
may also be a relevant target to be investigated
because they are based on the HHL algorithm. 
Finally, to extend the class of target channels, we need to develop a gadget for systematically decomposing a small part of the entire circuit, such as the 
Toffoli gate contained in the circuit for the factorization algorithm \cite{Oonishi-2022}.

\begin{acknowledgments}
We thank Hiroyuki Harada and Kaito Wada for helpful discussions.
This work was supported by MEXT Quantum Leap Flagship Program Grants No. JPMXS0118067285 and No. JPMXS0120319794.
\end{acknowledgments}

\appendix
\section{Gate decomposition of CNOT and Toffoli gates}
\label{app:gate-decomp}
Here we give the expressions of the gate decomposition of CNOT and Toffoli gates 
in Eqs.~(\ref{eq:CNOT}) and (\ref{eq:Toffoli}), respectively. 
The three-qubit QFT gate in Eq.~(\ref{eq:QFT}) is decomposed to 1524 channels, and 
thus we do not show those terms here.

CNOT gate can be decomposed as \cite{endo2018practical}
\begin{align}
[U^{(n)}]
&\,=\,
-\frac{1}{2}[IR_X]
+\frac{1}{2}[IX]
+[I\pi_X]
-\frac{1}{2}[R_ZI]
\nonumber\\
&
+[R_ZR_X]
-\frac{1}{2}[R_ZX]
+\frac{1}{2}[ZI]
-\frac{1}{2}[ZR_X]
\nonumber\\
&
+[ZX]
-[Z\pi_X]
+[\pi_ZI]
-[\pi_ZX]
\,,
\end{align}
where $[AB]$ means $[A\otimes B]$.
The corresponding overhead is $\gamma=\sum_i|c_i|=9$\,.

Toffoli gate can be decomposed as
\begin{align}
[U^{(n)}]
&\,=\,
\frac{3}{4} [IR_ZR_X]
+\frac{3}{4} [R_ZIR_X]
+\frac{1}{2} [I\pi_ZR_X]
+\frac{1}{2} [\pi_ZIR_X]
\nonumber\\
&
-\frac{1}{2} [IZR_X]
-\frac{1}{2} [ZIR_X]
-\frac{3}{4} [IIR_X]
+\frac{1}{2} [IR_Z\pi_X]
\nonumber\\
&
+\frac{1}{2} [R_ZI\pi_X]
+\frac{1}{2} [R_Z\pi_ZI]
+\frac{1}{2} [\pi_ZR_ZI]
-\frac{1}{2} [IR_ZX]
\nonumber\\
&
-\frac{1}{2} [R_ZIX]
-\frac{1}{2} [R_ZZI]
-\frac{1}{2} [ZR_ZI]
-\frac{3}{4} [IR_ZI]
\nonumber\\
&
-\frac{3}{4} [R_ZII]
+\frac{3}{4} [R_ZR_ZI]
-            [I\pi_Z\pi_X]
-            [\pi_ZI\pi_X]
\nonumber\\
&
+            [II\pi_X]
+            [I\pi_ZI]
+            [\pi_ZII]
-            [\pi_Z\pi_ZI]
\nonumber\\
&
-\frac{1}{2} [I\pi_ZX]
-\frac{1}{2} [\pi_ZIX]
+\frac{5}{8} [IZX]
+\frac{5}{8} [ZIX]
\nonumber\\
&
+\frac{3}{8} [IIX]
-\frac{1}{2} [IZ\pi_X]
-\frac{1}{2} [ZI\pi_X]
-\frac{1}{2} [Z\pi_ZI]
\nonumber\\
&
-\frac{1}{2} [\pi_ZZI]
+\frac{3}{8} [IZI]
+\frac{3}{8} [ZII]
+\frac{5}{8} [ZZI]
\nonumber\\
&
+\frac{3}{8} [III]
-            [R_Z\pi_ZR_X]
-            [\pi_ZR_ZR_X]
+\frac{1}{4} [R_ZZR_X]
\nonumber\\
&
+\frac{1}{4} [ZR_ZR_X]
+\frac{1}{2} [Z\pi_ZR_X]
+\frac{1}{2} [\pi_ZZR_X]
-\frac{1}{4} [ZZR_X]
\nonumber\\
&
-            [R_ZR_Z\pi_X]
+\frac{1}{2} [R_Z\pi_ZX]
+\frac{1}{2} [\pi_ZR_ZX]
-\frac{1}{4} [R_ZZX]
\nonumber\\
&
-\frac{1}{4} [ZR_ZX]
+\frac{1}{4} [R_ZR_ZX]
+\frac{1}{2} [R_ZZ\pi_X]
+\frac{1}{2} [ZR_Z\pi_X]
\nonumber\\
&
+            [\pi_Z\pi_ZX]
-            [Z\pi_ZX]
-            [\pi_ZZX]
+\frac{5}{8} [ZZX]
\nonumber\\
&
+            [Z\pi_Z\pi_X]
+            [\pi_ZZ\pi_X]
-            [ZZ\pi_X]
\,,
\end{align}
where $[ABC]$ means $[A\otimes B\otimes C]$.
The corresponding overhead is $\gamma=\sum_i|c_i|=37$\,.

\bibliography{draftNotes}
\bibliographystyle{apsrev4-1}

\end{document}